\documentclass[letter]{aa} 

\usepackage{graphicx}
\usepackage{txfonts}
\begin{document}

   \title{Revised spectroscopic parameters of SH$^+$ from ALMA\thanks{This 
paper makes use of the following ALMA data: ADS/JAO.ALMA$\#$2012.1.00352.S. 
ALMA is a partnership of ESO (representing its member states), NSF (USA) 
and NINS (Japan), together with NRC (Canada) and NSC and ASIAA (Taiwan), 
in cooperation with the Republic of Chile. The Joint ALMA Observatory is 
operated by ESO, AUI/NRAO and NAOJ.} and IRAM~30\,m\thanks{This 
paper  makes use of observations obtained with the IRAM 30\,m telescope. 
IRAM is supported by INSU/CNRS (France), MPG (Germany), and IGN (Spain).} 
observations}

   \author{Holger S.~P. M{\"u}ller
          \inst{1}
          \and
          Javier R. Goicoechea\inst{2}
          \and
          Jos\'e Cernicharo \inst{2}
          \and
          Marcelino Ag\'undez \inst{2}
          \and\\
          J\'er\^ome Pety \inst{3,4}
          \and
          Sara Cuadrado \inst{2,5}
          \and
          Maryvonne Gerin \inst{4}
          \and
          Gaelle Dumas \inst{3}
          \and 
          Edwige Chapillon \inst{3,6,7}
          }

\institute{I.~Physikalisches Institut, Universit\"at zu K\"oln, Z\"ulpicher Str. 77, 50937 K\"oln, Germany.\\
              \email{hspm@ph1.uni-koeln.de}
   \and   
Instituto de Ciencias de Materiales de Madrid (CSIC), 28049 Cantoblanco, Madrid, Spain.
   \and
IRAM, 300 Rue de la Piscine, 38406 Saint Martin d'H\`eres, France. 
   \and
CNRS UMR8112, LERMA, Observatoire de Paris, Ecole Normale Sup{\'e}rieure and 
Universit{\'e} Pierre et Marie Curie, 75014~Paris, France.
   \and
Centro de Astrobiolog\'{\i}a (CSIC-INTA), Carretera de Ajalvir km~4, 28850 Torrej\'on de Ardoz, Madrid, Spain.
   \and
CNRS, LAB, UMR5804, 33270 Floirac, France.
   \and
Universit{\'e} de Bordeaux, LAB, UMR5804, 33270 Floirac, France.
             }

   \date{Received 5 August 2014 / Accepted September 3 2014}

 
  \abstract{
Hydrides represent the first steps of interstellar chemistry. Sulfanylium (SH$^+$), 
in particular, is a key tracer of energetic processes. We used  ALMA and the IRAM~30\,m 
telescope to search for the lowest frequency rotational lines of SH$^+$ toward the Orion Bar, 
the prototypical photo-dissociation region illuminated by a strong UV radiation field. 
On the basis of previous \textit{Herschel}/HIFI observations of SH$^+$, 
we expected to detect emission of the two SH$^+$ hyperfine structure (HFS) components 
of the $N_J = 1_0 - 0_1$ fine structure (FS) component near 346\,GHz. While we did not 
observe any lines at the frequencies predicted from laboratory data, we detected 
two emission lines, each $\sim$15\,MHz above the SH$^+$ predictions and with 
relative intensities and HFS splitting expected for SH$^+$. The rest frequencies of 
the two newly detected lines are more compatible with the remainder of the SH$^+$ 
laboratory data than the single 
line measured in the laboratory near 346\,GHz and previously attributed to SH$^+$. 
Therefore, we assign these  new features to the two SH$^+$ HFS components of the 
$N_J = 1_0 - 0_1$ FS component and re-determine its spectroscopic parameters, which 
will be useful for future observations of SH$^+$, in particular if its lowest frequency 
FS components are studied. Our observations demonstrate the suitability of these 
lines for SH$^+$ searches at frequencies easily accessible from the ground.
    }

\keywords{ISM: individual objects: \object{Orion Bar} -- radio lines: ISM -- 
                 ISM: molecules -- molecular data -- line: identification}

\titlerunning{Re-determination of SH$^+$ spectroscopic parameters}
\authorrunning{M{\"u}ller, Goicoechea, Cernicharo et al.}

   \maketitle


\section{Introduction}

Hydrides, which consist of one heavy atom and one or more H atoms, are of fundamental 
importance in the interstellar medium (ISM). Unfortunately, their fundamental transitions 
at submillimeter (sub-mm) wavelengths are often difficult to observe from the ground. 
The Herschel Space Observatory and the Stratospheric Observatory for Infrared Astronomy 
(SOFIA) have increased our knowledge about hydrides in space considerably. 
Recent detections include ArH$^+$ with \textit{Herschel} 
\citep{ArH+_det_2013,ArH+_diff_2014} and SH with SOFIA \citep{SH_det_2012}.

The reactive molecular ion SH$^+$ (sulfanylium) is an interesting probe of energetic 
processes in the ISM. In particular, SH$^+$ is only detectable in significant amounts 
if the very high endothermicity (0.86\,eV or $\sim$9860\,K) of the gas-phase reaction 
\begin{equation}
\mbox{S$^+$ + H$_2$($\varv=0$) $\rightarrow$ SH$^+$(X$^3$$\Sigma^-$) + H} 
\end{equation}
can be overcome. Using the Atacama Pathfinder EXperiment (APEX) 12\,m telescope, 
\citet{Men11} discovered recently sub-mm SH$^+$ absorption lines in the envelope 
of the prolific star-forming region Sagittarius~B2(M), Sgr\,B2(M) for short, 
a very strong continuum source close to the Galactic center, and in the diffuse 
interstellar clouds on the line of sight toward this source. The ubiquity of SH$^+$ 
in diffuse clouds was re-emphasized soon thereafter by \textit{Herschel}/HIFI observations 
toward several Galactic sight lines by \citet{God12}. These authors proposed that the 
required energy for reaction (1) arises from turbulent dissipation, shocks, or shears 
in these very low density clouds. Sub-mm SH$^+$ emission lines were also detected by 
\textit{Herschel} in higher density environments such as the high-mass star-forming region 
W3\,IRS5 \citep{Ben10} or the Orion Bar photo-dissociation region (PDR) \citep{Nag13}. 
These are strongly UV-irradiated environments where the gas attains high temperatures 
($\lesssim 1000$\,K) and where H$_2$ molecules are UV-pumped to vibrationally 
excited states. In addition, in these PDR environments, reaction (1) becomes 
exothermic when H$_2$ molecules are in the $\varv=2$ or higher vibrational levels, 
thus enhancing the SH$^+$ abundance \citep[e.g.,][]{Zan13}. Finally, if doubly ionized 
sulfur atoms co-exist with H$_2$ molecules in X-ray dissociation regions (XDRs), 
such as Galactic center clouds \citep[e.g.,][]{God12,Etx13}, the very exothermic reaction 
\begin{equation}
\mbox{S$^{++}$ + H$_2$ $\rightarrow$ SH$^+$ + H$^+$ + \rm{8.79 eV}} 
\end{equation}
can be a significant source of SH$^+$ \citep[e.g.,][]{Abe08}.

The strongest ground state FS component of SH$^+$(X$^3 \Sigma^-$) at $\sim$526\,GHz 
(\mbox{$N_J = 1_2 - 0_1$}) cannot be observed from the ground because of the water in 
Earth's atmosphere. The $\sim$683\,GHz (\mbox{$N_J =1_1 - 0_1$}) and $\sim$893\,GHz 
(\mbox{$N_J =2_1 - 1_1$}) transitions can be observed from the ground, 
albeit with some difficulty. Transitions of the \mbox{$N_J = 1_0 -0_1$} FS component 
at $\sim$346\,GHz \citep{Sav04} lie in a  more accesible frequency window. 
Indeed, \citet{Men11} searched for the SH$^+$ $\sim$346, $\sim$683, and $\sim$893\,GHz 
absorption lines toward Sgr\,B2(M). 
Unfortunately, the $\sim$346\,GHz lines are heavily blended with absorption components 
of the nearby CO \mbox{$J=3-2$} line caused by Milky Way spiral arm molecular clouds. 
\citet{Sta07} detected an emission feature at 345930\,MHz in the massive protostar AFGL\,2591,  
but it may be caused by $^{34}$SO$_2$, a common species in these sources, so they could not 
draw firm conclusions about SH$^+$. \citet{Orion-Bar_1mm_2006} presented first the results 
of a 1\,mm line survey toward the Orion Bar with the APEX~12\,m telescope. They did not 
report the detection of SH$^+$. However, they observed the denser HCN clump \citep{Lis03} 
and not the most exposed UV-irradiated PDR gas layers where SH$^+$ is expected.

In this letter we present the first ground-based detection of SH$^+$ lines toward the Orion 
Bar CO$^+$ peak \citep[][and references therein]{Sto95,Goi11}. We show that the previously 
reported SH$^+$ \mbox{$N_J=1_0-0_1$} line frequencies at $\sim$346\,GHz are not correct 
and use the frequencies derived from our observations to compute improved SH$^+$ 
spectroscopic parameters and line frequencies.

\section{Observations}

\begin{figure}[t]
\centering
\includegraphics[scale=0.63,angle=0]{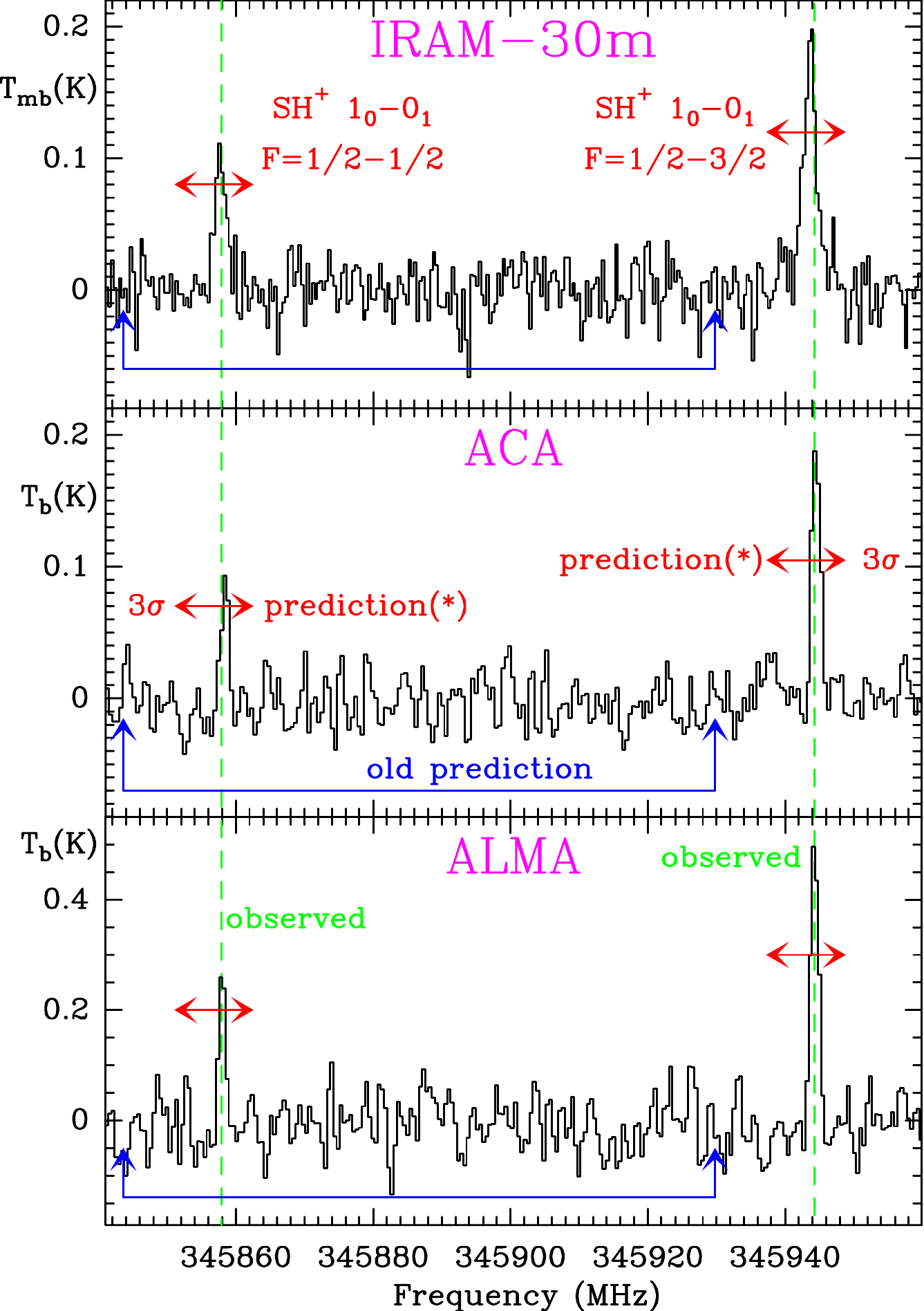}

\caption{From top to bottom: IRAM~30\,m, ACA, and ALMA spectra toward the Orion Bar between 
345.8\,GHz and 345.9\,GHz. Blue arrows show the predicted frequencies of the SH$^+$, 
$N_J = 1_0-0_1$, $F=1/2-1/2$ and $F=1/2-3/2$ lines before this work \citep[based on experiments 
by][]{Sav04}. Red arrows (prediction*) show the expected line frequencies for a Hamiltonian fit 
that excludes the 345929.8\,MHz line observed by \citet{Sav04} and attributed to SH$^+$; see 
Sect.~\ref{parameter-determination}.}
\label{Fig_observations}
\end{figure}

\subsection{Single-dish IRAM~30\,m observations}

\mbox{IRAM~30\,m} telescope observations were taken in 2013 using the EMIR330 receiver 
and the FFTS backend in the wide mode that covers a 16\,GHz instantaneous bandwidth per 
polarization at a channel spacing of 195\,kHz. They are part of a complete millimeter 
line survey toward the Orion Bar dissociation front \citep{Cua14} and include specific 
deep integration searches for SH$^+$ \mbox{$N_J=1_0-0_1$} lines. The target position is 
at \mbox{$\mathrm{\alpha_{2000}=05^{h}\,35^{m}\,20.8^{s}\,}$}, 
\mbox{$\mathrm{\delta_{2000}=-\,05^{\circ}25'17.0''}$}, close to the so-called 
CO$^{+}$ peak \citep{Sto95}. The observing procedure was position switching (PSW) 
with the reference position located at a \mbox{($-$600",0")} offset to avoid the 
extended molecular emission from the Orion Molecular Cloud. 
The antenna temperature, $T^{*}_{_{\rm A}}$, was converted to the main beam temperature, 
$T_{_{\rm MB}}$, using an antenna efficiency of 42\,\%. 
At this frequency, the IRAM~30\,m telescope provides an angular resolution of 7$''$. 
Figure~\ref{Fig_observations} (top panel) shows the resulting spectrum between 345.8\,GHz 
and 345.9\,GHz after baseline subtraction. The integration time was 2\,h, and the rms 
noise was $\sim$30\,mK per channel. The angular resolution and sensitivity of these 
observations are a factor of $\sim$2.5 better than those of \citet{Orion-Bar_1mm_2006}. 
Atmospheric calibration of the IRAM~30\,m data was carried out with the 
ATM program \citep[Cernicharo, J. 1985, IRAM internal report;][]{Pardo01}. 

\subsection{Interferometric ALMA and ACA observations}

Atacama Large Millimeter Array (ALMA) observations of the Orion Bar were carried out in 
2013 and 2014 during ALMA \textit{Cycle~1} observations. They belong to the 2012.1.00352.S 
project ``The fundamental structure of molecular cloud edges: from clumps to 
photoevaporation'' \mbox{(P.I.: J.R. Goicoechea)}. They consist of mosaicking 
observations covering a $\sim$40$''$$\times$40$''$ field centered  at 
\mbox{$\mathrm{\alpha_{2000}=05^{h}\,35^{m}\,20.6^{s}\,}$}, 
\mbox{$\mathrm{\delta_{2000}=-\,05^{\circ}25'20.0''}$}.

The ALMA array observations used twenty-seven 12\,m antennas. The field of view was observed 
as a mosaic of 27 Nyquist-sampled pointings in the C32-2 configuration, providing a 
typical resolution of $\sim$1$''$. Observations were carried out with the band\,7 receiver 
near 346\,GHz using the ALMA correlator backend at a spectral resolution of 488.3\,kHz 
($\sim$0.4\,km\,s$^{-1}$) over a total bandwidth of 937.5\,MHz. Atacama Compact Array (ACA) 
is used to complement ALMA observations of extended sources by producing short-spacing 
visibilities filtered out by the ALMA array. At the time of  observations, the ACA array 
used nine 7\,m antennas. The ACA mosaic contained ten pointings Nyquist-sampled fields and 
provided a typical resolution of $\sim$3.6$''$. The total observing time was about 6.5\,h 
and 2.2\,h for ACA and ALMA observations respectively. The data were calibrated within 
the CASA software package using the standard algorithms. The calibrated uv tables were 
then exported to GILDAS\footnote{See http://www.iram.fr/IRAMFR/GILDAS}, where the data 
were deconvolved using the H\"ogbom CLEAN algorithm. All technical details on the data 
reduction and the complete data set will be presented elsewhere. 
For the purpose of this work (showing the 345.8$-$345.9\,GHz window observed by different 
instrumentation) we only deconvolved the ACA and ALMA visibilities separately.


\section{Observational results}
\label{obs-res}

Figure~\ref{Fig_observations} shows, from top to bottom, the IRAM~30\,m, ACA, and ALMA 
spectra toward the Orion Bar dissociation front; we note that ACA and ALMA spectra are 
averaged over the same region of the dissociation front. The abscissa axis represents 
the rest frequency (in MHz) using a local standard of rest velocity (v$_{\rm LSR}$) of 
10.50\,km\,s$^{-1}$. We derived this v$_{\rm LSR}$ value from the accurately known 
HCO$^+$ \mbox{$J=4-3$} line frequency\footnote{We used \mbox{$\nu$(HCO$^+ 4-3$)} = 
356734.2244\,(9)\,MHz. All publicly available rotational lines up to $J=17$ and with 
uncertainties below 50\,kHz \citep{Tint07,Buf06,Hir08,Caz12} were used to compute HCO$^+$ 
line frequencies with MADEX \citep{Cer12}. The residual of the fit is 39\,kHz.} 
observed in the same region of the dissociation front. It agrees, within 
$\sim$0.2\,km\,s$^{-1}$, with typical velocity centroids of other reactive 
molecular ions observed with single-dish telescopes toward the Orion Bar 
\citep[see, e.g.,][]{Fue03,Orion-Bar_1mm_2006,Nag13}.

We expected to detect the SH$^+$ $F=1/2-1/2$ and $F=1/2-3/2$ HFS lines of the $N_J = 1_0-0_1$ 
FS component near 346\,GHz on the basis of the previous \mbox{\textit{Herschel}/HIFI} detection 
of the higher energy \mbox{SH$^+$ $N_J=1_2-0_1$} lines at $\sim$526\,GHz toward the Orion Bar 
\citep{Nag13}. Blue arrows in Fig.~\ref{Fig_observations} show the absence of emission 
at frequencies derived from previous laboratory data of \citet{Sav04} for all three telescopes. 
Our spectra, however, show two emission lines shifted by $\sim$15\,MHz ($\sim$13\,km\,s$^{-1}$) 
to higher frequencies with respect to the frequencies indicated by blue arrows, which could not 
be assigned to other carriers in the MADEX, CDMS \citep{CDMS_2001,CDMS_2005}, and JPL 
\citep{JPL-catalog_1998} spectroscopic catalogs. Table~\ref{table:1} shows results from 
Gaussian fits to the two lines, which were detected with up to 12\,$\sigma$ and 7\,$\sigma$ 
significance levels. Averaging the three observations, we obtain 345944.35\,MHz and 
345858.27\,MHz for the two lines. Given the complex gas kinematics in Orion, implying an 
uncertainty in v$_{\rm LSR}$, we adopted a 1\,$\sigma$ systematic uncertainty of 0.2\,MHz for 
the rest frequency determination. We note that the line near 345.9~GHz is only $\sim$61\,MHz 
($\sim$53\,km\,s$^{-1}$) above the CO $J=3-2$ line frequency.


\section{New SH$^+$ spectroscopic parameters}
\label{parameter-determination}

The two unpaired electrons ($S = 1$) of sulfanylium lead to a $^3 \Sigma ^-$ electronic 
ground state which causes all rotational levels $N$ to be split into three with $J = N$, 
$N \pm 1$, except for the $N = 0$ level for which only $J = 1$ exists. The lowest order 
FS parameters are $\lambda$ and $\gamma$ which describe the (electron) spin-spin coupling 
and the (electron) spin-rotation coupling. Rotational (or centrifugal distortion) and 
vibrational corrections, the latter for excited vibrational states, may appear in the 
Hamiltonian. The proton magnetic moment ($I_{\rm H} = 1/2$) splits every FS level into 
two HFS levels. The main HFS parameters here are $b_F$ and $c$, the scalar and tensorial 
electron spin-nuclear spin coupling parameters.

The selection rules $\Delta F = 0, \pm1$ hold strictly. The spin-conserving transitions 
($\Delta F = \Delta J = \Delta N$) are the strongest ones at higher quantum numbers. 
Other transitions have non-negligible intensities, in particular at low-$N$. 
The $N = 1 - 0$ ground state rotational transition is split into three FS components 
$J = 0 - 1$, $2 - 1$, and $1 - 1$ near 346, 526, and 683\,GHz, respectively, which are 
split further into 2, 3, and 4 HFS components. An energy level diagram is shown, e.g., 
in \citet{Men11}.

The rotation-vibration spectrum of SH$^+$ was recorded up to $\varv = 2 - 1$ \citep{SH+_IR_1986} 
and $\varv = 4 - 3$ \citep{SH+_IR_1989}. \citet{SH+_LMR_1987} investigated the rotational 
spectrum of SH$^+$ in its ground and first excited vibrational state between 0.51\,THz and 
1.62\,THz by laser magnetic resonance (LMR). More recently, \citet{Sav04} employed source- 
and velocity-modulation to determine accurate rest frequencies of all three HFS 
components near 526\,GHz of the $N = 1 - 0$ transition of SH$^+$ as well as one of two 
HFS components near 346\,GHz. In each of these four studies, spectroscopic parameters were 
determined only from their own data. In contrast, \citet{SH+_parameters_2009} 
used the field-dependent \citep{SH+_LMR_1987} and field-free \citep{Sav04} rotational 
data as well as the rovibrational data of the fundamental ($\varv = 1 - 0$) band 
\citep{SH+_IR_1986,SH+_IR_1989} to determine ground state spectroscopic parameters 
as well as vibrational corrections if they were needed. One of the present authors 
(HSPM) used a slightly different approach to create an entry for the CDMS catalog because 
the SPFIT program \citep{spfit_1991} does not permit the use of field-dependent data. 
The LMR transition frequencies extrapolated to zero field by \citet{SH+_parameters_2009} 
were fit together with the field-free rotational data \citep{Sav04} and all of the 
rovibrational data \citep{SH+_IR_1986,SH+_IR_1989} to determine Dunham-type parameters 
$P_{i,j}$ for SH$^+$, with $i$ and $j$ indicating vibrational and rotational corrections, 
respectively.


\begin{table}
\begin{center}
\caption{Spectroscopic parameters$^{a}$ (MHz, cm$^{-1}$) of sulfanylium, SH$^+$.}
\label{spec-parameter}
\begin{tabular}[t]{lr@{}l}
\hline \hline
Parameter                     & \multicolumn{2}{c}{Value} \\
\hline
$Y_{10}$$^b$                  &    2\,547&.4948~(104) \\
$Y_{20}$$^b$                  &     $-$49&.4293~(90)  \\
$Y_{30}$$^b$                  &         0&.2097~(30)  \\
$Y_{40} \times 10^3$$^b$      &     $-$16&.01~(34)    \\
$Y_{01}$                      &  278\,094&.99~(36)    \\
$Y_{11}$                      & $-$8\,577&.33~(85)    \\
$Y_{21}$                      &        16&.15~(32)    \\
$Y_{02}$                      &     $-$14&.7380~(76)  \\
$Y_{12} \times 10^3$          &       122&.9~(27)     \\
$Y_{03} \times 10^3$          &         0&.46$^c$     \\
$\lambda _{00}$               &  171\,488&.3~(58)     \\
$\lambda _{10}$               &    $-$471&.8~(146)    \\
$\lambda _{20}$               &     $-$79&.1~(67)     \\
$\lambda _{01}$               &      $-$1&.13~(24)    \\
$\gamma _{00}$                & $-$5\,036&.29~(91)    \\
$\gamma _{10}$                &       116&.4~(20)     \\
$\gamma _{20}$                &         3&.52~(64)    \\
$\gamma _{01}$                &         0&.432~(35)   \\
$b_{F,0}(^1$H)                &     $-$56&.884~(81)   \\
$b_{F,1}(^1$H$)-b_{F,0}(^1$H) &      $-$3&.46~(79)    \\
$c(^1$H)                      &        33&.60~(67)    \\

\hline
\end{tabular}\\[2pt]
\end{center}
\footnotesize{
$^a$ Numbers in parentheses are one standard deviation in units of the least significant digits.\\
$^b$ In units of cm$^{-1}$.\\
$^c$ Kept fixed to the value derived by \citet{SH+_parameters_2009}.\\
}
\end{table}


\citet{Sav04} observed all three HFS components of the $N_J = 1_2 - 0_1$ transition of 
sulfanylium (around 526.1\,GHz) between 0.2\,MHz and 0.5\,MHz higher than calculated from 
the parameters of the LMR study \citep{SH+_LMR_1987}; the deviations are well within the 
uncertainties of $\sim$1.6\,MHz from that study. These rest frequencies improve the 
accuracies of the major HFS parameters $b_F$ and $c$, as well as the origin of the FS 
component, which depends on several rotational and fine structure parameters. 
The $F = 1/2 - 3/2$ HFS component was found by \citet{Sav04} only 1.23\,MHz higher than 
calculated from parameters in \citet{SH+_LMR_1987}, well within the uncertainty of 
$\sim$18\,MHz. The uncertainty is much larger because this FS component was not accessed 
in the LMR study. The associated $F = 1/2 - 1/2$ HFS component, weaker by a factor of two, 
was not detected despite significant signal averaging. Its frequency, $\sim$86\,MHz lower 
than the $F = 1/2 - 3/2$ HFS component, is well constrained by the frequency of this 
HFS component and from the data near 526.1\,GHz.

As outlined in section~\ref{obs-res}, no emission lines were detected in the Orion Bar 
at the frequencies expected from \citet{Sav04}. However, the two emission lines, 
each observed $\sim$15\,MHz higher, had the expected 1\,:\,2 intensity ratio 
and spacing required to assign them to the $N_J = 1_2 - 0_1$ FS component. 
In order to test the feasibility of these assignments, we omitted the laboratory 
transition frequency near 346\,GHz from the data set which was used to create the 
CDMS catalog entry. Interestingly, the rms error of the fit improved significantly 
from 0.898 to 0.826. More importantly, the two HFS components were now predicted at 
345856.8\,MHz and at 345943.0\,MHz, very close to our observed emission lines 
(Fig.~\ref{Fig_observations}) and each with predicted 1\,$\sigma$ uncertainties of 
3.9\,MHz. These uncertainties are considerably smaller than those from \citet{SH+_LMR_1987}, 
mainly because of the extensive rovibrational transition frequencies used in the present fit 
and because of the remaining data from \citet{Sav04}. Hence, we used averages over the 
three independent observations given in section~\ref{obs-res} to redetermine the 
SH$^+$ spectroscopic parameters. The resulting parameters are given in 
Table~\ref{spec-parameter}. The rms error of this fit is 0.820, demonstrating that 
our transition frequencies are more compatible with the remainder of the laboratory 
data than the single frequency near 346\,GHz determined by \citet{Sav04}. 
Updated predictions of the rotational spectrum of SH$^+$ will be available in the CDMS 
catalog.\footnote{Internet address: http://www.astro.uni-koeln.de/cdms/catalog} 
Data up to 1\,THz are given in Table~A2.



Errors in rest frequencies determined in the laboratory or through astronomical 
observations are known to occur especially in sparse data sets. Recent error correction 
of data pertaining to astrophysically important hydrides include the correction 
of the fundamental transition of H$_2$D$^+$ \citep{H2D+_HD2+_rot_2008} and of 
the fundamental transitions of CH$^+$ isotopologues \citep{CH+_isos_rot_2010}.

The changes in spectroscopic parameters are small for the most part and occur predominantly 
in rotational and FS parameters. The most significant changes of 3\,$\sigma$ occur in 
$Y_{01} \approx B_e$ and $Y_{02} \approx -D_e$. The largest change in magnitude occurs 
in $\lambda _{00}$, but corresponds only to $\sim$1\,$\sigma$. The strong, spin-conserving 
transitions are only slightly affected at lower frequencies. In addition, the transitions 
of the \mbox{$N_J =1_1 - 0_1$} FS component ($\sim$683\,GHz) are now predicted 2\,MHz lower 
than in the first SH$^+$ CDMS catalog entry. Changes from the predictions by 
\citet{SH+_parameters_2009} are a bit more complex, but deviate on average by $\sim$2\,MHz. 
Larger deviations of several megahertz may occur for transitions with flip of the electron 
spin (e.g., at $\sim$893\,GHz) or at higher frequencies.

The 345929.8\,MHz line observed in the laboratory, and attributed to \mbox{SH$^+$, 
$N_J=1_0-0_1$}, $F=1/2-3/2$ by \citet{Sav04}, is not detected at the sensitivity limit 
of our observations. Here we demonstrate that the 345929.8\,MHz line is not SH$^+$, and 
we improve the SH$^+$ spectroscopic parameters with the detection of the 345944 and 
345858\,MHz lines, which arise from the edge of the Orion Bar, consistent with the 
expected SH$^+$ formation route in PDRs, the reaction of S$^+$ with vibrationally 
excited H$_2$ \citep[e.g.,][]{Nag13,Zan13}. Nevertheless, it may still be useful 
to reinvestigate the $N = 1 - 0$ transition of SH$^+$ in the laboratory using methods 
such as those by \citet{Sav04,H2D+_HD2+_rot_2008,CH+_isos_rot_2010,C3H+_rot_2014}, 
in particular in combination with a study of $^{34}$SH$^+$.



\begin{acknowledgements}

We thank the Spanish MINECO for funding support under grants CSD2009-00038, AYA2009-07304, 
and AYA2012-32032. We thank the European Research Council for funding support under ERC-2013-Syg 
610256-NANOCOSMOS. HSPM is supported by the German Bundesministerium f{\"u}r Bildung und 
Forschung (BMBF) via the ALMA Regional Center (ARC) Node project 5A11PK3 for maintanance 
and upgrade of the CDMS. JRG thanks the Observatoire de Paris/ENS and IRAM~(Grenoble) 
where part of this work was carried out. SC was supported by a FPI-INTA grant.

\end{acknowledgements}



\begin{appendix}

\section{Complementary tables}


\begin{table*}[t]
\caption{Gaussian fits to the $F = 0.5 - 1.5$ and $0.5 - 0.5$ lines of the $N_J = 1_0 - 0_1$ 
         FS component of SH$^+$ observed with three different telescopes toward the Orion Bar 
         (for v$_{\rm LSR}$=10.50\,km\,s$^{-1}$).}
\label{table:1}
\centering
\begin{tabular}{c c c c c c c}
\hline\hline
          &             & Rest frequency   & $\displaystyle{\int} T_{_{\mathrm{\bf mb}}}d$v & Line width & \\ 
Telescope &  Line       &  (MHz)           &  (mK\,km\,s$^{-1}$)  &  (km\,s$^{-1}$) & Signal-to-noise \\ 
\hline
IRAM~30\,m&             &                  &                        &                   &          \\
          & $1.5 - 0.5$ & 345944.22\,(20)  &  383\,(30)             &  1.95\,(19)       & $\sim$10 \\
          & $0.5 - 0.5$ & 345858.38\,(20)  &  185\,(31)             &  1.68\,(32)       & $\sim$5  \\
ACA       &             &                  &                        &                   &          \\
          & $1.5 - 0.5$ & 345944.55\,(20)  &  260\,(20)             &  1.23\,(10)       & $\sim$12 \\
          & $0.5 - 0.5$ & 345858.42\,(20)  &  114\,(25)             &  1.10\,(26)       & $\sim$6  \\
ALMA      &             &                  &                        &                   &          \\
          & $1.5 - 0.5$ & 345944.29\,(20)  &  668\,(51)             &  1.20\,(10)       & $\sim$12 \\
          & $0.5 - 0.5$ & 345858.00\,(20)  &  295\,(58)             &  0.92\,(20)       & $\sim$7  \\
\hline
\end{tabular}
\tablefoot{Parentheses show the Gaussian fit uncertainties (in units of the least significant digits). 
For the fitted frequencies, we adopt an overall \mbox{1$\sigma$ uncertainty} of 0.2\,MHz ($\sim$0.2\,km\,s$^{-1}$).
This should be considered as a systematic uncertainty due to uncertainties in v$_{\rm LSR}$ (see text). 
IRAM~30\,m spectrum smoothed to $\sim$0.4\,km\,s$^{-1}$, similar to ALMA and ACA; Signal-to-noise 
determined on the peaks and refering to $\sim$0.4\,km\,s$^{-1}$ resolution for IRAM~30\,m and to the native 
spectral resolution for ALMA and ACA. 
}
\end{table*}



\begin{table*}
\caption{Quantum numbers, frequencies (MHz), uncertainties unc. (MHz), Einstein $A$ values 
         (10$^{-4}$\,s$^{-1}$), upper $g_{\rm up}$ and lower $g_{\rm lo}$ state degeneracies, 
         and upper $E_{\rm up}$ and lower $E_{\rm lo}$ state energies (cm$^{-1}$) of 
         sulfanylium, SH$^+$, below 1\,THz.}
\label{spectrum_to_1THz}
\begin{center}
\begin{tabular}{cccr@{}lr@{}lr@{}lccr@{}lr@{}l}
\hline\hline
$N' - N''$ & $J' - J''$ & $F' - F''$ & \multicolumn{2}{c}{Frequency} & \multicolumn{2}{c}{unc.} 
 & \multicolumn{2}{c}{$A$} & $g_{\rm up}$ & $g_{\rm lo}$ & \multicolumn{2}{c}{$E_{\rm up}$} 
& \multicolumn{2}{c}{$E_{\rm lo}$} \\[1pt]
\hline

$1-0$ & $0-1$ & $0.5-0.5$ & 345858&.1957 & 0&.1494 &  1&.15 & 2 & 2 & 11&.5395 &  0&.0029 \\
$1-0$ & $0-1$ & $0.5-1.5$ & 345944&.4205 & 0&.1496 &  2&.30 & 2 & 4 & 11&.5395 &  0&.0000 \\
$1-0$ & $2-1$ & $1.5-0.5$ & 526038&.7348 & 0&.0699 &  8&.00 & 4 & 2 & 17&.5496 &  0&.0029 \\
$1-0$ & $2-1$ & $2.5-1.5$ & 526047&.9440 & 0&.0714 &  9&.59 & 6 & 4 & 17&.5471 &  0&.0000 \\
$1-0$ & $2-1$ & $1.5-1.5$ & 526124&.9597 & 0&.0723 &  1&.60 & 4 & 4 & 17&.5496 &  0&.0000 \\
$1-0$ & $1-1$ & $1.5-0.5$ & 683334&.0618 & 0&.5139 &  2&.90 & 4 & 2 & 22&.7964 &  0&.0029 \\
$1-0$ & $1-1$ & $0.5-0.5$ & 683359&.9250 & 0&.4790 & 11&.59 & 2 & 2 & 22&.7973 &  0&.0029 \\
$1-0$ & $1-1$ & $1.5-1.5$ & 683420&.2867 & 0&.4972 & 14&.48 & 4 & 4 & 22&.7964 &  0&.0000 \\
$1-0$ & $1-1$ & $0.5-1.5$ & 683446&.1499 & 0&.4942 &  5&.79 & 2 & 4 & 22&.7973 &  0&.0000 \\
$2-1$ & $1-1$ & $0.5-0.5$ & 893065&.8800 & 0&.9514 & 19&.77 & 2 & 2 & 52&.5868 & 22&.7973 \\
$2-1$ & $1-1$ & $0.5-1.5$ & 893091&.7433 & 0&.9763 &  9&.89 & 2 & 4 & 52&.5868 & 22&.7964 \\
$2-1$ & $1-1$ & $1.5-0.5$ & 893126&.2353 & 1&.0505 &  4&.94 & 4 & 2 & 52&.5888 & 22&.7973 \\
$2-1$ & $1-1$ & $1.5-1.5$ & 893152&.0985 & 0&.9475 & 24&.73 & 4 & 4 & 52&.5888 & 22&.7964 \\

\hline
\end{tabular}\\[2pt]
\end{center}

\end{table*}

\end{appendix}


\end{document}